\pdfoutput=1
\documentclass[conference]{IEEEtran}
\usepackage[a4paper, total={6in, 8.5in}]{geometry}
\usepackage{float}
\usepackage{array}
\usepackage{graphicx}
\usepackage{gensymb}

\graphicspath{{./images/}}
\DeclareGraphicsExtensions{.jpg}

\begin{document}

\title{Intelligent Sensors and Monitoring System for Low-cost Phototherapy Light for Jaundice Treatment}

\author{\IEEEauthorblockN{Paul M. Cabacungan}
\IEEEauthorblockA{{\textit{Ateneo de Manila University}} \\
Quezon City, Philippines \\
pcabacungan@ateneo.edu}\\

\IEEEauthorblockN{Dr. Gregory L. Tangonan}
\IEEEauthorblockA{{\textit{Ateneo de Manila University}} \\
Quezon City, Philippines \\
gtangonan@ateneo.edu}

\and
\IEEEauthorblockN{Carlos M. Oppus}
\IEEEauthorblockA{{\textit{Ateneo de Manila University}} \\
Quezon City, Philippines \\
coppus@ateneo.edu}\\

\IEEEauthorblockN{Ivan B. Culaba}
\IEEEauthorblockA{{\textit{Ateneo de Manila University}} \\
Quezon City, Philippines \\
iculaba@ateneo.edu}

\and
\IEEEauthorblockN{Dr. Jeremie E. De Guzman}
\IEEEauthorblockA{{\textit{Ateneo de Manila University}} \\
Quezon City, Philippines \\
jedeguzman@ateneo.edu}\\

\IEEEauthorblockN{Nerissa G. Cabacungan}
\IEEEauthorblockA{{\textit{Ateneo de Manila University}} \\
Quezon City, Philippines \\
ncabacungan@ateneo.edu}}

\maketitle

\begin{abstract}
A prototype of a low-cost phototherapy light system (LPLS) was deployed by the Ateneo Innovation Center (AIC) at a public hospital in Metro Manila, Philippines.  It underwent clinical investigation for two years under the supervision of licensed physicians in a public tertiary hospital.

This paper presents the process of upgrading the LPLS in order to enhance capabilities and improve efficiency yet remain affordable. The following features were added: (1) a visual and auditory monitoring system in order to remotely oversee the infant from the nurse station; (2) an automation system that stores data about the device's light intensity and bulb temperature and records ambient humidity; (3) an alarm system that activates the warning lights if sensor readings are in critical level and if the bulbs need to be replaced; and (4) a time setting to manually set the time of operation and automatically turn-off the device as programmed

The upgrades increased the system's cost but it remained cheaper than the ones commercially available.  For deployment in remote or off-grid hospitals, the system was equipped with a solar-powering provision.
\end{abstract}
\IEEEoverridecommandlockouts
\section{Introduction}
\IEEEPARstart{N}{eonatal} jaundice is a condition in infants characterized by the presence of high level of bilirubin.  Increased levels of serum bilirubin cause irreversible brain damage and kernicterus in infants. Jaundice occurs in the first two weeks of life in 25\% to 60\% of full-term newborns, and in 80\% of preterm newborns [1].

Treatment for jaundice was discovered serendipitously in 1957 by an observant nursing sister in United Kingdom. The nun reported that the infants whose crib was located in the nursery area exposed to sunlight had their jaundice fade [2].  Exposing the baby directly to sunlight is still being practiced.  There are disadvantages to this practice: harmful intensity of sunlight rays that can cause acute erythema or sunburn, risk of dehydration, and ultraviolet ray exposure that can cause skin damage, skin cancer and eye damage, among others [3][4][5].

A low-cost phototherapy design that filters UV from sunlight was recently published [6].  However, sunlight is not always available during rainy days or in the evening. If present, it is harmful to the skin for most time of the day.  Thus, this design failed to optimize sun exposure and treatment.

Phototherapy remains the most frequently used treatment when serum bilirubin levels exceed physiological limits. The most important intervention for infants with severe hyperbilirubinemia is immediate exposure to phototherapy.  It converts bilirubin into water soluble photo-products that can bypass the hepatic conjugating system and be excreted without further metabolism. The clinical response to phototherapy depends on the efficacy of the phototherapy device, as well as the infant's rates of bilirubin production and elimination [7].

Phototherapy can be delivered using several types of conventional light sources, including daylight, white or blue fluorescent bulbs and filtered halogen bulbs.  The efficacy and ability of these light sources to provide intensive phototherapy vary widely.  Some light sources cannot be too close to the infant due to high heat production and unstable broad wavelength light output. In recent years, a new type of light source, light-emitting diodes (LEDs), has been developed and studied as possible light sources for the phototherapy. LEDs are power efficient, portable, with low heat production, not heavy and have a longer life span [2].  Blue LEDs emit a high intensity narrow band of blue light overlapping the peak spectrum of bilirubin breakdown resulting in potentially shorter treatment time.  These characteristics of LEDs make them an optimal light sources for a phototherapy device [8].

Phototherapy, despite being the recommended standard of treatment for hyperbilirubinemia, is prohibitive due to its high cost and expensive replacement bulbs.  In the United States, a second-hand phototherapy light unit costs roughly US\$2,000 and a brand new unit is US\$3,000.  The treatment cost ranged from US\$800 to US\$1,000 depending on the severity of the infant's condition.  In the Philippines, a brand-new phototherapy unit can cost PhP 150,000 (approximately US\$3,000) depending on the quality and efficiency of the unit.  The treatment is similarly costly.  For instance, in one private provincial hospital in the Philippines, treatment costs around \$16/day. It is likely that not all hospitals and health units have ready access to phototherapy units financially and logistically.  Where it is available in public settings, anecdotes point to too few units available causing patients to share units and a long queue for using the device.  This in turn leads to sub-optimal exposure time and poorer outcomes.

Varied designs of phototherapy light units have been tried out to address concern with cost and portability, among others.

A portable and low-cost phototherapy blanket uses 200 blue LED light bulbs to provide light therapy to jaundiced infants. Its flexible photovoltaic panel harnesses solar energy to power the blanket's battery, eliminating the need for electrical power sources [9].

The Brezinski's Bili-Hut phototherapy design is a battery-powered pop-up tent lined with LEDs. It is collapsible, highly portable, can fit into a shipping tube yet big enough for a newborn to fit inside comfortably. The LEDs that line its inner surface are arranged in a radial array, and can shine the right kind of blue light at the right intensity over a baby's entire body.  All the parts are off-the-shelf. The interior lining, for example, is made from a material used for hydroponic gardening. This design can run off 12-volt power [10].

A light-up pajama for jaundice treatment uses polymer optical fibers to weave and bend at just the right angle around perpendicularly arranged polyester threads. When connected to LEDs, this generates the right amount of light, emitting it throughout the fabric with constant intensity [11][12].  However, Waltz (2017) cited that wearable medical devices cannot compete with conventional phototherapy light units because they were found very difficult to be certified by regulators [11].

The above-mentioned designs are all imported.  Thus, busted bulbs may not be easily replaceable and local technicians might not be able to repair them. In this light, the AIC made a prototype that is locally designed and less costly - the low cost phototherapy light system or LPLS.  However, there are some concerns that emerged during its clinical trial.

First, the light intensity of light-emitting diodes (LEDs) deteriorated over time. This was seen in the phototherapy prototype designed and deployed by AIC in a public hospital in Metro Manila. The readings reflected a decrease in irradiance from 84 to 59.40 microwatt per square centimeter per nanometer, which was around 30\% decrease, after a three-year span [13].  Similarly, the wavelength decreased from 462.1 nm to 457.15 nm which is still within the acceptable range.

Another concern was that phototherapy light units are mainly run by electricity, making it dependent on the stability of electrical lines. When power supply is intermittent or cut-off, the phototherapy is disabled.

AIC locally designed a more efficient phototherapy light unit with the intention to address the above concerns.  The upgraded design is affordable, with components that are readily available in the local market, and with technical design that is understandable for local technicians to trouble shoot and repair.

\section{Methodology}
This is a design and engineering study and hence does not require prior ethical approval or ethics clearance. This study was partially funded by Ateneo de Manila University, where the authors are affiliated with, so there is no conflict of interest to be declared.

We conceptualized the design of the low-cost phototherapy light system (LPLS) shown in Fig. 1.

We identified the components to be used, learned about operating them then canvassed and compared the prices and efficiency. We chose and purchased all equipment and materials needed and tested if everything was functioning well: LED bulbs, solar panel, batteries, charge controllers, DC-AC converters, and electrical cable. In fabricating the LPLS, we used 20 pieces of 3-watt LED blue bulb, one 10-watt computer fan with an AC to DC power adaptor, roller base, adjustable iron stand, fuse, wires and switches. We conducted a stress test by running it continuously for 72 hours and we monitored and recorded the wavelength, irradiance, and power consumption.

\begin{figure}[h]
\centering
\includegraphics[width=3.19 in, height=2 in]{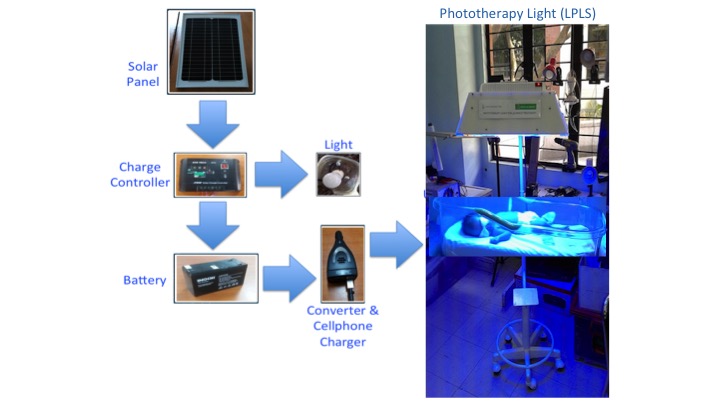}
\label{block_diagram}
\caption{Block diagram of the Low-cost Phototherapy Light System.}
\end{figure}

We made another unit to integrate intelligent sensor and monitoring system. We called this upgraded prototype as the Improved Low-cost Phototherapy Light System or ILPLS.

\section{Improved Low-cost Phothotherapy Light System}
The ILPLS was designed to have the same functions as the initial prototype. It was made to meet the clinical needs using low-cost and locally available components. The main components were the eight pieces of 7-watt LED blue spotlight, which are more robust and have the same cost as the 3-watt bulbs. The wavelength was within the 400 - 500 nanometers (peak at 460nm) which are specifically used for administering phototherapy [14]. The measured irradiance was three times stronger than that of the 3-watt bulb. The design of an 8-bulb unit exceeded the total irradiance of the previous unit. The bulbs were positioned alongside each other at a distance of 13 centimeters apart (from center to center). This engineering design made the light beams overlap by 43\% resulting to a more intensified lighting. There is an 11\% intersection of light beams coming from the bulbs positioned across each other, which are 22 centimeters apart. The 22-centimeter distance was an engineering trade-off to give way to the fans that provide fail-safe mechanism. Fig. 2 shows the LED array.

\begin{figure}[h]
\centering
\includegraphics[width=3in, height=1.75in]{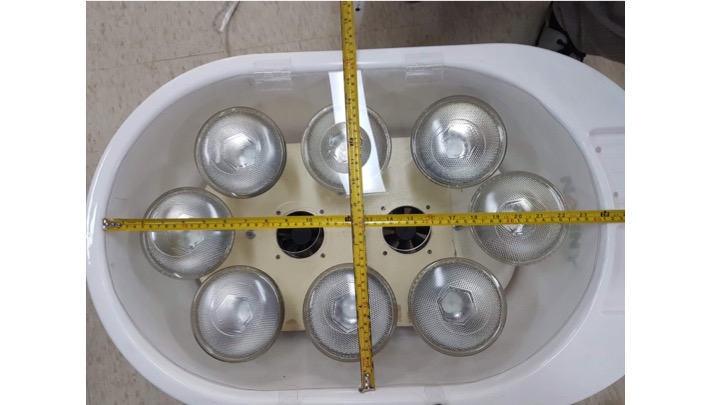}
\label{LED_array}
\caption{LED array with two ventilation fans.}
\end{figure}

Despite the distance, light intensity in the area covered by overlapping beams still fell within the range of required irradiance for jaundice treatment.

We also purposely designed the LED array so the irradiation patterns overlapped in the exposure area covering 150\% of the 45-centimeter average infant length. Light intensity was determined with a photodetector that was calibrated using the RS PRO ISM410. We made sure that the average intensity exceeded the minimum needed by an engineering margin of 300\% to accommodate any changes in performance.

These bulbs were enclosed in a plastic basin with an acrylic cover. We also installed two units of 3-watt computer fans (12Vdc) to increase the fail-safe mechanism. We put a camera that can give visual monitoring to baby's condition and video records it and a microphone that can remotely monitor infant's audio or voice at the nurse station. We assembled the intelligent system and developed the codes to be used for the microcontrollers, sensors and actuators. We calibrated and tested them for accuracy. The monitoring system will report, if the irradiation levels diminish too much, necessitating an LED replacement. We made a provision such that the system can accept pre-set values for ambient temperature. Afterwards, the intelligent system's PCB layout and housing were designed and made. The whole system was supported by a galvanized iron stand with an adjustable height and movable base.

We tested the unit for durability by running it continuously for five days, non-stop, and no significant issues were encountered. The block diagram of ILPLS is shown in Fig. 3.

\begin{figure}[h]
\centering
\includegraphics[width=2.75in, height=2in]{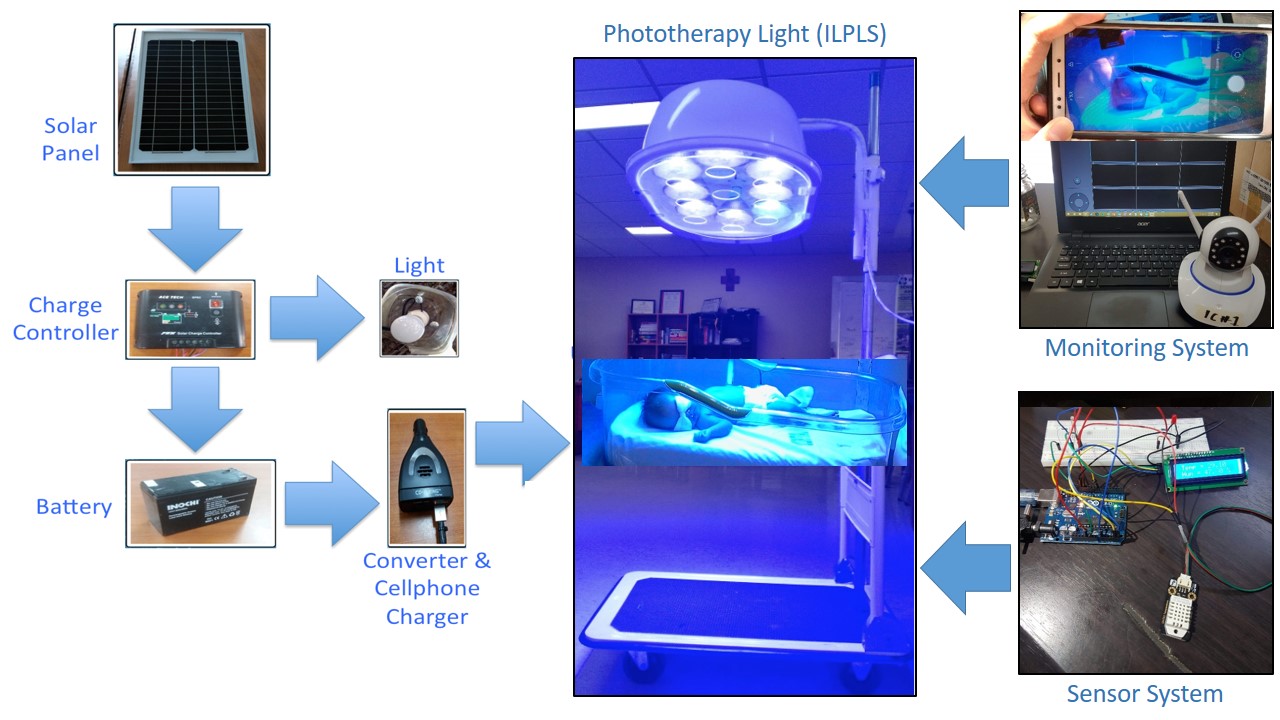}
\label{Block_diagram_ILPLS}
\caption{Block diagram of the improved low-cost phototherapy light system.}
\end{figure}

\section{Results and Discussion}
We had successfully wired an intelligent system for ILPLS unit using light-dependent resistors (LDRs), temperature and humidity sensors, microcontrollers, microphone, camera and internet connection.  The microphone and camera remotely monitor the infant's audio and visually monitor the baby's condition.    We devised a way to maintain and keep things in check through the calibrated photosensor for the light intensity, temperature sensor of the light bulbs and the ambient humidity. The light intensity sensor is used to monitor the LED intensity which can detect deterioration or damage to the LEDs.  The intelligent sensors can also detect surrounding temperature and humidity and trigger alarm system when the readings go beyond the set threshold.    

A commercial spectrometer was used to measure the right wavelength of the LEDs.  Using Ocean Optics Model USB 4000, we were able to obtain 20.61 nm (437.38 nm to 457.68 nm) wavelength at full width half maximum (FWHM) and peak wavelength at 447.68 nanometers.

The ILPLS, like the LPLS and the commercial phototherapy unit, emitted more than the required minimum irradiance of 30 microwatt/square-centimeter/nanometer for effective treatment.  Using RS PRO ISM 410 in measuring irradiance of the three different units (Table 1), we can see that the ILPLS has the strongest irradiance.  Notice that its irradiance is three times higher than the required minimum value.  This is to ensure that the light intensity, even if it would degrade 30\% in three years (like what was observed during clinical trial), would remain twice the required strength. 
LPLS and ILPLS are also more energy-efficient, which makes provision of solar powering less costly and more feasible. 

In Table 1, we presented the differences of the three phototherapy units.

We were able to run the ILPLS for 10 hours powered by two 50-ampere-hour commercial lead-acid car battery that were pre-charged by two 50-watt solar panels. In order to run the unit for 24 hours, five car batteries or 250-ampere-hour ampacity and five solar panels would be needed. We can implement a battery swapping system to ensure continuous operation of the unit. It would entail charging of five batteries while the other five batteries are in use. This solar photovoltaic set-up using a ten-battery swapping system costs about \$1,250. Aside from solar-powering, a pedal-powered battery charging could also be used and its cost is almost the same. This alternative power source where one pedals a bicycle attached to an alternator can charge batteries even at night time. The additional cost for these alternative power sources enables deployment of the phototherapy system in off-grid hospitals and adds value of treatment accessibility to infants with jaundice who live in far flung areas without electrical service.

\begin{table}[h]
\caption{COMPARISON OF PROTOTYPES AND A COMMERCIAL UNIT}
\centering
\begin{tabular}{|>{\centering\arraybackslash} m{0.55in}|>{\centering\arraybackslash} m{0.55in}|>{\centering\arraybackslash} m{0.55in}|>{\centering\arraybackslash} m{0.55in}|}
\hline
\textbf{Parameters} & \textbf{LPLS} & \textbf{ILPLS} & \textbf{Commercial Unit} \\
\hline
\textbf{Light Bulb} & Twenty 3-watt LED & Eight 7-watt LED & Five 20-watt Fluorescent \\
\hline
\textbf{Wavelength} & 462.1 nm to 476.4 nm & 437.8 nm to 458.48 nm  & (5-white, 2-blue) \\
\hline
\textbf{Irradiance at 30 cm Distance} & 84 $\mu$W/cm2/nm & 90 $\mu$W/cm2/nm & 70 $\mu$W/cm2/nm \\
\hline
\textbf{Total Power} & 75 watts & 90 watts & 400 watts \\
\hline
\textbf{Ventilation } & One  fan & Two  fans & Air-cooled \\
\hline
\textbf{Unit Cost} & PhP 20,000 & PhP 35,000 & PhP 150K \\
\hline
\textbf{Sensor \& Monitoring System} & None & Yes & None \\
\hline
\end{tabular}

\end{table}

During the stress test of continuous 120-hour operation, the ILPLS bulbs' temperature ranged from 35.8\degree°C to 39.0\degree°C, which is an acceptable range, at an ambient average temperature of 32.5\degree°C. No issues were encountered at all.

The upgrades made the ILPLS more expensive than the initial prototype but remained significantly cheaper than the commercial unit. Having been designed and fabricated locally, availability of spare parts is almost always assured. Manual of operation comes with the unit to guide the user for its operation, trouble shoot and repair.

\section{Conclusion and Recommendations}
The study of Bagunu and Perez (2018) revealed the effectiveness of the first prototype of the LPLS in treating fifty (50) infants with jaundice.  Their study revealed that after 24 hours of exposure,  adjusted mean bilirubin levels of those who underwent LED phototherapy was significantly lower as compared to conventional phototherapy (17.27 $\pm $ 2.31 mg/dl versus 18.64 $\pm $ 2.81  mg/dl, p=0.000). After 48 hours, there was consistently lower bilirubin in the experimental group (13.37 $\pm $ 2.213 mg/dl versus 15.4 $\pm $ 2.496 mg/dl, p=0.000) and there were no reported side effects in both groups.  They concluded that use of LED phototherapy showed significantly better bilirubin-lowering effect.  The improved prototype (ILPLS), which was similar to the first one but used LED spotlights that emitted greater irradiance, could offer faster treatment.  The system's heat production was found within tolerable level so as not to pose any risks on the infant.  

The intelligent sensor and monitoring system was successfully integrated into the phototherapy light system without affecting the unit's optimal functioning. The camera system is intended to provide digital records of the patient ID, time stamp, unit tag, audio tags by medical personnel, location of the phototherapy that can readily interface with nursing station and hospital medical records.  The smart camera was able to transmit images and sounds within the 10-meter radius in a laboratory setting, despite the presence of walls, demonstrating that data collection at nearby stations is possible. This approach is scalable and useful to situations where multiple phototherapy units are operated at the same time, the data of the ongoing phototherapy can get recorded easily, serving as a complement to the nurses records. Four photosensors or light-dependent resistors (LDRs) enabled provision of more accurate light intensity readings, after determining ambient lighting that may interfere with the phototherapy light intensity values. The two-in-one sensor also indicated temperature and humidity. The cameras can be used as a backup intensity monitor, thus application is under development. All readings were stored in a system SD card.  The timer setting and alarm system worked as programmed.

The careful selection of spare parts made ILPLS 75\% cheaper than the commercial unit despite the additional features that were included.  Alternative power sources opened possible deployment of the unit to off-grid communities.  The mock set-up of the ILPLS is found in Figure 4.

\begin{figure}[h]
\centering
\includegraphics[width=1.75 in, height=1.75 in]{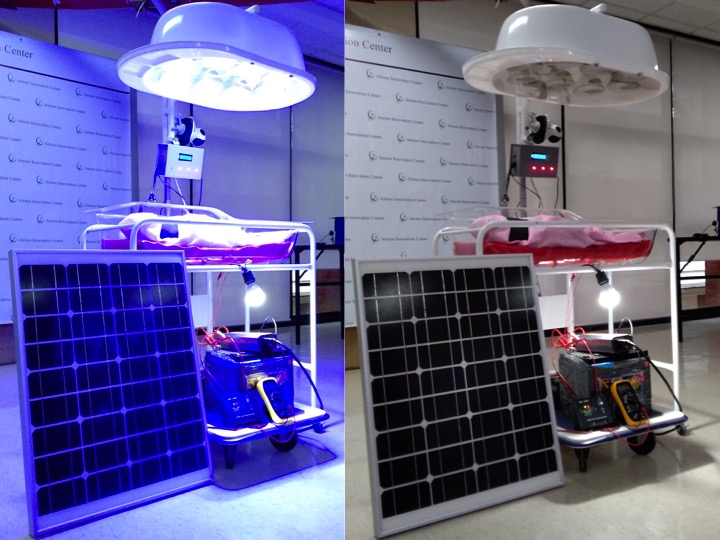}
\label{Setup}
\caption{Mock set-up of the improved low-cost phototherapy light system.}
\end{figure}

We recommend that deployment of ILPLS in hospitals can be done to find out its actual performance and further test its durability. The efficacy of the intelligent sensors and monitoring system can also be investigated on the field and the ease of use can be looked into. Further research can be done to explore ways to prevent or slow down weakening of LED lights' intensity and wavelength over time.  Consultation and discussion with medical practitioners can provide valuable input for its further development.

We also recommend cross-calibration of the Silicon photodiodes used in monitoring with this calibrated detector, including the Webcam Silicon photodiodes. A simple application averages the light intensity over groups of pixels and this output is another intensity monitor. This way we can gather the average values of the exposure from over the infant's body. Each unit is scanned for performance and calibrated against the reference photometer. 

Given the importance of having a digital record of treatment for patient records, hospital operations, and PhilHealth payment, we anticipate the further IT that will enhance the system usefulness without adding extra cost.  In an expanded prototype, using a camera, we can use image recognition to record a patient label, capture images of infant's posture during treatment measure, total exposure, nurse attendant and LED readout at the unit.   

Calibrated open source apps in hand-held devices can also be used to detect the following:  (1) target color of the infant's skin that will indicate improvement in health condition, (2) irradiance of light to know if the bulbs are functioning properly for optimal healing capacity, (3) temperature inside the system to determine the conducive heat level for the infant.  A smart phone can be used to access these apps and an LCD to display the readings. There is more room for technical upgrade of ILPLS. Yet, as it is, it already offers greater accessibility to jaundice treatment in an affordable way with more helpful features and an alternative power source for off-grid health facilities.

\section*{Acknowledgment}
Two phototherapy prototypes built by AIC went under extensive research acceptance tests in public hospital in Metro Manila by our collaborators -- Dr. Vanessa Marie V. Calabia, Dr. Viel M. Bagunu and Dr. Ma. Lucila M. Perez. We thank them for their efforts and valuable feedback.

We are grateful to our researchers, Mr. Reymond P. Cao, Mr. Kerwin G. Caballas, Mr. Lawrence R. Ibarrientos, and the AIC interns and OJTs who worked in AIC for the past five years. In their capacities, they gave their valuable assistance in the different stages of fabricating the prototype.

We are indebted to the University Research Council (URC) of the Ateneo de Manila University for granting financial support to come up with this paper and to explore further developments of the ILPLS.

Our gratitude goes to the Ateneo Innovation Center for serving as a hub for ILPLS’ development since its initial conceptualization until its actual deployments.

\end{document}